\documentclass[twocolumn,showpacs,preprintnumbers,amsmath,amssymb,showkeys,pra]{revtex4}

\usepackage{graphicx}% Include figure files
\usepackage{dcolumn}% Align table columns on decimal point
\usepackage{bm}% bold math
\usepackage{amsmath}
\usepackage{amsfonts}
\usepackage{amssymb}
\usepackage{color}

\begin{document}

\title{Discrete Family of Dissipative Soliton Pairs in Mode-Locked Fiber Lasers}
\author{Aleksandr Zavyalov}
\email{aleksandr.zavyalov@uni-jena.de} \affiliation{Institute of
Condensed Matter Theory and Solid State Optics,
Friedrich-Schiller-Universit\"{a}t Jena, Max-Wien-Platz 1, 07743
Jena, Germany}
\author{Rumen Iliew}
\affiliation{Institute of Condensed Matter Theory and Solid State
Optics, Friedrich-Schiller-Universit\"{a}t Jena, Max-Wien-Platz 1,
07743 Jena, Germany}
\author{Oleg Egorov}
\affiliation{Institute of Condensed Matter Theory and Solid State
Optics, Friedrich-Schiller-Universit\"{a}t Jena, Max-Wien-Platz 1,
07743 Jena, Germany}
\author{Falk Lederer}
\affiliation{Institute of Condensed Matter Theory and Solid State
Optics, Friedrich-Schiller-Universit\"{a}t Jena, Max-Wien-Platz 1,
07743 Jena, Germany}

\date{\today}

\begin{abstract}
We numerically investigate the formation of soliton pairs (bound states) in
mode-locked fiber ring lasers. In the distributed model (complex cubic-quintic
Ginzburg-Landau equation) we observe a discrete family of soliton pairs with
equidistantly increasing peak separation. This family was identified by two
alternative numerical schemes and the bound state instability was disclosed by
a linear stability analysis. Moreover, similar families of unstable bound state
solutions have been found in a more realistic lumped laser model with an
idealized saturable absorber (instantaneous response). We show that a
stabilization of these bound states can be achieved when the finite relaxation
time of the saturable absorber is taken into account. The domain of stability
can be controlled by varying this relaxation time.
\end{abstract}


\maketitle

\section{\label{sec:level1}Introduction}

Since the experimental observation of stable solitary pulses (SPs) in optical
fibers \cite{1} their interactions attracted a great deal of interest. In the
early stages the propagation of light pulses was described using the nonlinear
Schr\"{o}dinger equation (NLSE). Relying on a perturbation approach it was
shown that SP interactions depend on the relative phase between them: in-phase
solitons attract each other, while out-of-phase solitons repel \cite{2, 3, 4,
5}, and for an arbitrary phase the interaction appears more involved \cite{5}.

For a more realistic description of real systems different
perturbations were added to the NLSE, accounting for losses
\cite{6}, intrapulse Raman scattering \cite{7, 8, 9}, third-order
dispersion \cite{7, 8} or spectral filtering. For the description of
a complete communication line with a periodic arrangement of signal
regeneration (semiconductor amplifier, saturable absorber) and
propagation in the fiber (effects as nonlinearity and absorption) a
distributed description with the so-called complex cubic-quintic
Ginzburg-Landau equation (CQGLE) was established \cite{11}. Another
optical system of primary importance, which is commonly modeled by
this equation, is the mode-locked fiber ring laser, because it
essentially consists of the same optical components in a closed loop
\cite{12}. Despite its success in the description of mode-locked
lasers, the assumptions used to derive the CQGLE are not always
fulfilled in reality.

In this contribution we investigate the formation of bound states in fiber
lasers exploiting the CQGLE and a more realistic lumped model. From exact
numerical simulations of the CQGLE we obtain a discrete family of BS stationary
solutions with different (equidistant) peak separation. Moreover, we show that
a lumped laser model which accounts additionally for finite temporal effects of
the absorber can have profound consequences for the BS stability.

The CQGLE can be considered as a perturbed NLSE where the complex coefficients
of dispersion and nonlinearity account for spectral filtering and saturation,
respectively, and a linear loss/gain is added. For certain parameter sets this
equation was shown to support solitary wave solutions, called dissipative
solitons, as well as bound states (BS) of pairs of SPs  \cite{13, 14, 15, 16,
17, 18, 19, 20, 21}. Careful numerical investigations \cite{16} have
demonstrated BS formation from a pair of initially resting SPs. Depending on
the dispersion regime and the initial phase difference the BS profile can vary.
Furthermore it was shown \cite{17, 18} that out-of-phase and in-phase SPs can
form unstable stationary BS solutions, because they represent saddle points in
the phase plane. In the framework of the CQGLE stable BSs were discovered by
Akhmediev et al. \cite{21} as stable two-soliton solutions with a $\pi/2$ phase
difference of the peaks. They were observed for anomalous dispersion only.

In an early work \cite{14} it was analytically shown that these dissipative
equations exhibit a discrete family of two-soliton solutions with different
(equidistant) peak separation. This analysis was based on both a perturbation
approach for the NLS equation with small pumping and dissipative terms and the
CQGLE with weak anomalous dispersion. To the best of our knowledge, however,
such discrete families of BSs have not yet been found numerically or
experimentally.

In the first section of this paper we briefly present the equations used for
modeling the fiber ring laser for different approximations. We depart from the
more general lumped description, where all individual elements in the cavity
are modeled by separate equations and the pulse is sequentially propagated
through these elements in the ring cavity. Then, we derive the CQGLE from the
lumped model by averaging over the full cavity. We obtain definite relations
between the coefficients of both models which is essential for the comparison
with experiments.

The second section is devoted to the discrete family of stationary BS solutions
derived from the CQGLE. In the numerical simulations we observe a discrete
family of stationary two-soliton solutions with different peak-to-peak
separation which we identify as different BS levels. These results are obtained
by means of two numerical schemes, viz.,  the solution of either the evolution
or the stationary problem. Based on the results of the stationary analysis, a
linear stability analysis is carried out where the instability growth rates
perfectly coincide with the results of the propagation model.

For a more realistic description we take advantage of the lumped laser model in
the third part, where the individual elements of the cavity are treated
separately, and show that the different BS levels can be stabilized by
accounting for a noninstantaneous, but very fast, response of the saturable
absorber (SA), whereas for an instantaneous absorber response the BSs remain
unstable. In the latter case the SA is described by the commonly used
instantaneous response approximation (ideal SA).

Interestingly, by varying the SA response time and saturation level the BS
solutions can be forced to change from stable to unstable behavior. Moreover,
all these stable two-pulse solutions exhibit a multilevel nature and are
located on a certain side of the phase plane depending on the level.

In what follows we are restricting ourselves to the \emph{normal
dispersion regime} because this is the realm of the so-called
all-normal-dispersion lasers, which attract a considerable deal of
interest presently. Compared to other fiber lasers they have a
simpler setup and allow for achieving higher pulse energies.
Energies as high as 256 nJ per pulse were achieved in a
large-mode-area photonic crystal fiber laser \cite{22}, and more
than 20 nJ for a usual single-mode fiber laser \cite{23}. Pulses
generated in all-normal-dispersion lasers depend nontrivially on the
interplay of spectral filtering and self-phase-modulation \cite{24}.

\section{\label{sec:level1}Models}

For the numerical modeling we use a simple scheme of a ring fiber laser, which
consists of a doped fiber, a saturable absorber and an output coupler. This
laser model allows for including the dominant effects into the simulations and
is still close to reality. In the lumped model the propagation through each
element is treated separately.

The propagation along the doped fiber is described by the modified nonlinear
Schr\"{o}dinger equation which is given by \cite{7, 25, 26} (when the carrier
optical frequency equals the dopant's atomic resonance frequency)

\begin{eqnarray} \begin{array}{p{1.5cm}p{1.0cm}p{1.5cm}}
\multicolumn{2}{l}{\dfrac{\partial U(z,t)}{\partial
z}+\dfrac{i}{2}(\beta_2+ig(z)T^2_1)\dfrac{\partial^2
U(z,t)}{\partial^2 t}} \\
&\multicolumn{2}{r}{=\dfrac{g(z)U(z,t)}{2}+i\gamma|U(z,t)|^2U(z,t)},
\end{array} \end{eqnarray}

\noindent where $U(z,t)$ is the envelope of the pulse, $z$ is the
propagation coordinate, $t$ is the retarded time, $\beta_2$ is the
second-order dispersion (GVD) coefficient, and $\gamma$ represents
the fiber nonlinearity. $g(z)$ is the saturable gain of the doped
fiber and $T_1$ is the dipole relaxation time (inverse linewidth of
the parabolic gain). Assuming that the conditions are close to
stationarity, the gain can be approximated by \cite{27}

\begin{equation}
g(z)=\frac{g_0}{1+\int\limits_\mathrm{pulse}|U(z,t)|^2dt/E^\mathrm{Gain}_\mathrm{sat}},
\end{equation}

\noindent where $g_0$ is the small-signal gain, which is defined by the pumping
level, and $E^\mathrm{Gain}_\mathrm{sat}$ is the saturation energy.

To describe the time-dependent semiconductor SA response we use the
Agrawal/Olsson model \cite{28}, which we term noninstantaneous SA response and
is given by

\begin{equation}
\begin{array}{lll}
\dfrac{\partial U(z,t)}{\partial z} &=&
-\dfrac{1}{2}\delta(z,t)U(z,t)
\\ \dfrac{\partial \delta(z,t)}{\partial t} &=& \dfrac{\delta_0 -
\delta(z,t)}{T_\mathrm{relax}} -
\dfrac{\delta(z,t)|U(z,t)|^2}{E^\mathrm{SA}_\mathrm{sat}},
\end{array}
\end{equation}

\noindent where $\delta(z,t)$ is the loss introduced by the absorber,
$\delta_0$ is the small-signal loss, $T_\mathrm{relax}$ is the recovery time
and $E^\mathrm{SA}_\mathrm{sat}$ is the saturation energy. For
$T_\mathrm{relax}<T_\mathrm{pulse}$ and taking into account that standard
absorbers are thin, we obtain  the well-known transmission equation for the SA
in the instantaneous response approximation (ideal SA) from system (3)

\begin{equation}
U_\mathrm{out}(t)=U_\mathrm{in}(t)\exp\left( -\frac{1}{2} \frac{\delta_0\Delta
z}{1+|U_\mathrm{in}(t)|^2/P_\mathrm{sat}} \right),
\end{equation}

\noindent where $\Delta z$ is the length of the SA, $\delta_0\Delta
z$ defines the modulation depth of the absorber and the saturation
power is defined as
$P_\mathrm{sat}=E^\mathrm{SA}_\mathrm{sat}/T_\mathrm{relax}$.

This lumped model describes the experimental setup appropriately but
it is too involved for analytical studies, including the linear
stability analysis. In order to obtain a single equation with
constant coefficients the periodic system is approximated by
averaging over the full cavity length using the guiding-center
soliton technique \cite{29, 30}, which relies on the field
decomposition into a product containing a periodic part and an
average amplitude. This model is justified if physically relevant
changes appear upon several round-trips. For the sake of simplicity
of the resulting equation the instantaneous response approximation
for the SA (4) is used and the saturation of the fiber gain is
neglected, i.e.\ we simplify Eq.\ (2) to $g(z)\approx g_0$ in the
limit $\int\limits_\mathrm{pulse}|U(z,t)|^2dt \ll
E^\mathrm{Gain}_\mathrm{sat}$. Eventually we obtain a single
evolution equation, which is a modified complex Ginzburg-Landau
equation (MGLE) given by

\begin{equation} \begin{array}{p{1.5cm}p{1.0cm}p{1.5cm}}
\multicolumn{2}{l}{i\dfrac{\partial V}{\partial
Z}+\dfrac{D}{2}\dfrac{\partial^2 V}{\partial \tau^2} + |V|^2V} \\
&\multicolumn{2}{r}{= i\theta V+i\beta\dfrac{\partial^2 V}{\partial \tau^2}-
\dfrac{i\rho V}{1+|V|^2/P^\mathrm{aver}_\mathrm{sat}}},
\end{array} \end{equation}

\noindent where $V$ is the normalized envelope of the pulse, $Z=z/L_\mathrm{D}$
is the normalized averaged propagation coordinate,
$L_\mathrm{D}=T^2_0/|\beta_2|$ is the dispersion length, $\tau=t/T_0$ is the
time normalized by the pulse duration $T_0$ and the other parameters are
defined as

\begin{equation}
\begin{array}{l}
D=-\mathrm{sgn}(\beta_2), \\ \beta=\dfrac{g_0 T^2_2}{2}|\beta_2|, \\
\rho=\dfrac{L_\mathrm{D} \delta_0 \Delta z}{2L_\mathrm{f}}, \\
\theta=L_\mathrm{D} \dfrac{g_0-k/L_\mathrm{f}}{2}, \\
P^\mathrm{aver}_\mathrm{sat}=P_\mathrm{sat}L_\mathrm{D}\gamma
\dfrac{1-\exp(-g_0 L_\mathrm{f})}{g_0 L_\mathrm{f}},
\\ V=U\left(L_\mathrm{D} \gamma \dfrac{\exp(g_0 L_\mathrm{f})-1}{g_0L_\mathrm{f}}\right)^{1/2},
\end{array}
\end{equation}

\noindent where $L_\mathrm{f}$ is the fiber length, which is assumed
to be equal to the cavity length and $k$ is the output loss. A
further simplification can be achieved by a Taylor expansion of the
last term of Eq.\ (4) up to second order in
$|V|^2/P^\mathrm{aver}_\mathrm{sat}$. Ultimately we obtain the
established CQGLE given by

\begin{equation} \begin{array}{p{1.5cm}p{1.0cm}p{1.5cm}}
\multicolumn{2}{l}{i\dfrac{\partial V}{\partial
Z}+\dfrac{D}{2}\dfrac{\partial^2
V}{\partial \tau^2} + |V|^2V} \\
&\multicolumn{2}{r}{= i\delta V+i\beta\dfrac{\partial^2}{\partial \tau^2} +
i\varepsilon |V|^2 V - i\mu |V|^4 V}, \end{array}
\end{equation}

\noindent where the linear and nonlinear gain/loss coefficients are
defined as

\begin{equation}
\delta = \theta - \rho,~ \varepsilon =
\dfrac{\rho}{P^\mathrm{aver}_\mathrm{sat}}, ~ \mu =
\dfrac{\rho}{\left(P^\mathrm{aver}_\mathrm{sat}\right)^2}.
\end{equation}

\noindent Hence, we can unambiguously relate the coefficients of the CQGLE and
MGLE to the coefficients of the lumped model which reflect the experimental
laser parameters.

\section{\label{sec:level1}The CQGLE model}

\subsection{\label{sec:level2}Bound state solutions}

To get a general picture of the two-SP interactions in the fiber
laser we start our considerations from the distributed laser model,
CQGLE (7). The reason is twofold, first most previous papers rely on
this model and second, it is more appropriate for identifying the
stationary solutions and applying the linear stability analysis.

From previous investigations \cite{13, 14, 15, 16, 17, 18, 19, 20,
21} it is known that BS formation is caused by linear and nonlinear
dissipative effects (right hand side of Eq.\ (7)). Thus to simplify
the analysis the number of system parameters was reduced from four
$(\delta,\beta,\varepsilon,\mu)$ to two $(\beta,k)$ such as
\cite{16}:

\begin{equation}
\delta = -k\beta/3,~ \varepsilon = \beta/2 + k\beta, ~ \mu =
\left(5/8\right)k\beta,
\end{equation}

\noindent where the parameter $k$ characterizes the deviation from
the particular point $(k=0)$ where arbitrary-amplitude solutions
exist \cite{31}.

We are working in the normal dispersion regime $D=-1$ with relatively strong
spectral filtering $\beta=1$. To be sufficiently far from the singularity we
choose $k=0.5$ where stable SPs exist. As initial conditions we use an in-phase
superposition of two resting pulses separated by a certain distance. In the
framework of the CGQLE the initial pulse profile is a critical issue with
respect to BS formation. Thus for a better convergence we use as input two SPs,
which are numerically obtained stationary one-soliton solutions for the
respective parameter set \cite{17}.

To get a more complete picture of two-pulse solutions we consider
the evolution for a wider range of initial separations and longer
propagation distance than in previous works, see Fig.\ 1. We observe
stationary BS formation beyond a certain distance (similar to
\cite{17}) and its succeeding destabilization, i.e. fusion of the
pulses (below we discuss this instability in more detail). For an
initial separation exceeding a critical value the SPs repel each
other, was also observed in \cite{16, 17}. However, a more careful
study discloses that this separation is no ordinary repulsion but
the formation of another BS with a larger peak separation (see Fig.\
1). In previous works it was usually assumed that the interaction
becomes too weak for such large peak separations and, hence, this
scenario was not thoroughly investigated. The evolution to the
second BS level has a much slower dynamics than that for the first
one and needs an approximately 50 times longer propagation distance
to form the stationary solution. This difference arises from the
weaker SP interaction. Except for the larger formation distance and
the slower formation dynamics there is no principal difference
between these two BS levels. Similarly, beyond another critical
initial separation we have observed the evolution towards a third BS
level. Unfortunately, these calculations are too time-consuming, but
intuitively it is clear that even higher order BS levels should
occur. Evidently, it may be anticipated that the dynamics becomes
progressively slower for higher levels and that all of them are
unstable.

\begin{figure}
\centerline{
\includegraphics[width=8.5cm, clip]{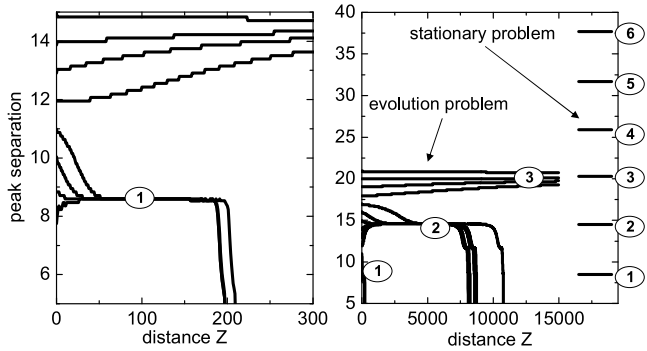} } \caption{Evolution of the separation for initially in-phase pulses
with different initial separation. The levels of stationary in-phase
BS solutions are displayed on the right axis on the right side. The
peak separation is given in units of $\tau$; model: CQGLE (7);
parameters: $k=0.5$, $\beta=1$.}
\end{figure}

To double-check these expectations and to identify arbitrary BS
levels, we solve the stationary problem using the relaxation and
Newton's methods in combination. Figure 1 (right) shows results
obtained from both the propagation and from the stationary analysis,
which are in perfect agreement in the intermediate section where the
propagation is stationary. The solution of the stationary problem
yielded six levels of in-phase BSs with different peak separation.
It is interesting to see that the peak separation $S$ changes from
level to level by a constant value (approximately $\Delta S=6$ for
the current set of parameters) and obeys the simple equation
$S^{n}_\mathrm{in-phase}\approx 2.5+6n$ where $n=1,2,3,...$
designates the BS level.

Also the first six levels of the out-of-phase stationary BS solutions were
found. Similarly, the dependence of the peak separation on the level (except
the first one) can be approximated by
$S^{n}_\mathrm{out-of-phase}\approx5.5+6n$, where $n=2,3,4,...$, while for the
first level $S^{1}_\mathrm{out-of-phase}=4.7$. We guess that this peculiarity
of the first level is caused by the strong SP overlap resulting in enhanced
nonlinear effects.

We note that for a $\pi/2$ phase difference between the pulses
neither stable nor unstable BS solutions have been found in the
\emph{normal dispersion regime}.

To understand the multilevel nature we consider the BS intensity and
phase profiles for the first three levels, shown in Fig.\ 2. From
this figure we may draw two conclusions. First, the intensity
profile of in-phase (out-of-phase) BS looks always the same,
independently of the level, and has a characteristic small dip in
the center. Second, there is a definite phase relation between the
BS constituents and between the levels. For example, for in-phase BS
levels the phase difference between the phase maximum and the phase
minimum (at the center) is a multiple of $\pi$ and can be written as
$\Delta \varphi^n_\mathrm{in-phase}\approx \pi n$, where
$n=1,2,3,...$ is the BS level. For out-of-phase solutions this phase
difference can be approximated by $\Delta
\varphi^n_\mathrm{out-of-phase}\approx 0.8+ \pi n$, where
$n=2,3,4,...$, while for the first level $\Delta
\varphi^1_\mathrm{out-of-phase}=3.6$.

\begin{figure}
\centerline{
\includegraphics[width=8.5cm, clip]{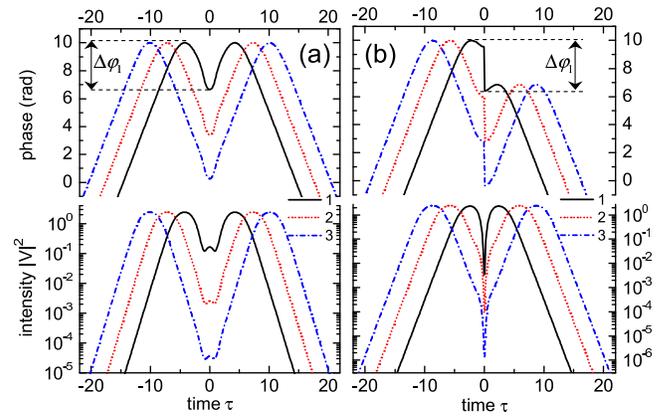} } \caption{(Color online) Intensity and phase profiles for the first three levels of
the (a) in-phase and (b) out-of-phase stationary BS solutions;
model: CQGLE (7); parameters: $k=0.5$, $\beta=1$.}
\end{figure}

Physically, the equidistant multilevel nature of the BSs can be
understood as constructive interference between SPs and can be
clearly seen from the almost linearly growing phase of the tails
(Fig.\ 2). On the other side, our results are in qualitative
agreement with Malomed's analysis \cite{14, 18} where the
oscillating tails of the SPs (linear phase change) evoke a periodic
interaction potential against the pulse separation, in spite of the
fact that this analysis \cite{14, 18} is based on a NLSE
perturbation approach and is valid just for the anomalous dispersion
regime. Moreover, we propose the magnitude $\Delta \varphi$ as a
convenient parameter for the definition of the BS level in
simulations or experiments.

We visualize the evolution of both pulses in a phase plane
$\left(\rho(z),\phi(z) \right)$ where $\rho(z)$ is the peak
separation and $\phi(z)$ their relative phase (difference of the
phase between peaks), see Fig.\ 3. In this plot we analyze the
evolution trajectories of two pulses for two relevant cases. In the
first case we consider two out-of-phase SPs with an initial
separation of $S=7$ which are evolving towards the first level
out-of-phase stationary BS solution. In the second case the
evolution of a pair of in-phase SPs with a larger initial separation
of $S=16.1$ (evolving towards the second level in-phase stationary
BS solution) is displayed. The evolution trajectory for the first
case, which may be partly described by a circle indicated by '1' in
Fig.\ 3, is well studied \cite{17}. The radius of this circle equals
the distance between the peaks of the first level out-of-phase BS.

\begin{figure}
\centerline{
\includegraphics[width=6cm, clip]{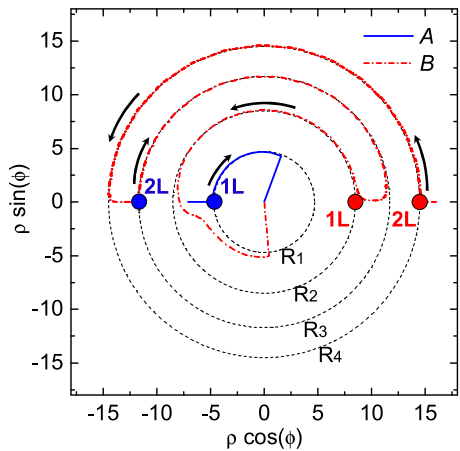} } \caption{(Color online) Soliton trajectories in
the phase plane for: two initially out-of-phase solitons with
separation 7 (line \textit{A}) and two initially in-phase solitons
with separation 16.1 (line \textit{B}). 1L - first level stationary
in-phase (red right) and out-of-phase (blue left) bound state
solution, 2L - second level stationary in-phase (red right) and
out-of-phase (blue left) bound state solution; model: CQGLE (7);
parameters: $k=0.5$, $\beta=1$.}
\end{figure}

To date, it was not shown that in the second case, where the initial
separation between the solitons is 16.1, the evolution trajectory
partly consists of four circles. From the plot (Fig.\ 3) we can
recognize that each circle corresponds to a certain BS. The radius
of the biggest one equals the distance between pulses of the second
level in-phase BS, being in agreement with the previous case. The
other circles '3', '2' and '1' correspond to the second level
out-of-phase, first level in-phase and first level out-of-phase
stationary BS solutions, respectively.

It is interesting to note that the evolution trajectory of two in-phase or
out-of-phase SPs attains simple geometrical forms. Moreover, during the
evolution they pass through all possible in-phase and out-of-phase stationary
BS solutions with ever smaller peak separation.

This can be explained by the earlier finding that the phase
difference between the pulses depends on their separation as a
consequence of interference phenomena between them.

\subsection{\label{sec:level2}Linear stability analysis}

In previous papers \cite{17, 18, 21} it was shown that in-phase and
out-of-phase first level BS are unstable. From our propagation
simulations we anticipate that the corresponding higher level BSs
are unstable as well. In order to confirm this hypothesis and to
provide a more complete picture we carry out a linear stability
analysis of the higher BS levels. We start with the first level
in-phase stationary solution. Its formation, the typical instability
behavior and final fusion of the two-pulse excitation are shown in
Fig.\ 4. It is interesting to note that the instability exhibits a
temporal asymmetry in the absolutely symmetric CQGLE.

\begin{figure}
\centerline{
\includegraphics[width=6cm, clip]{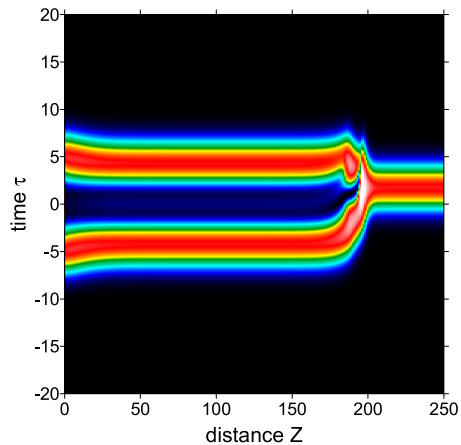} } \caption{(Color online) Bound state instability.
Two in-phase solitons form a
quasi-stationary bound state but fuse asymmetrically eventually;
model: CQGLE (7); parameters: $k=0.5$, $\beta=1$.}
\end{figure}

To perform the linear stability analysis we perturb the stationary
solution $V_0(t)\exp(ik_0z)$ as $V(z,t)=V_0(t)\exp(ik_0z)+\delta
V(z,t) = \left[V_0(t) + f\exp(\lambda z) + g\exp(\lambda^* z)
\right]\exp(ik_0z)$, where $V_0(t)$ was numerically calculated and
$f,g$ are the small amplitudes of the perturbation. The substitution
of this ansatz into the CQGLE (7) and its respective linearization
in $f,g$ leads to a system of linear, homogeneous equations. The
solvability condition provides the eigenvalues $\lambda$.
Eigenvalues with positive real part indicate that the perturbation
grows upon propagation and the stationary solution destabilizes. In
Fig.\ 5 the calculated eigenvalues for the unstable first level
in-phase BS are displayed and compared with the eigenvalues of the
stable SP solution.

\begin{figure}
\centerline{
\includegraphics[width=8.5cm, clip]{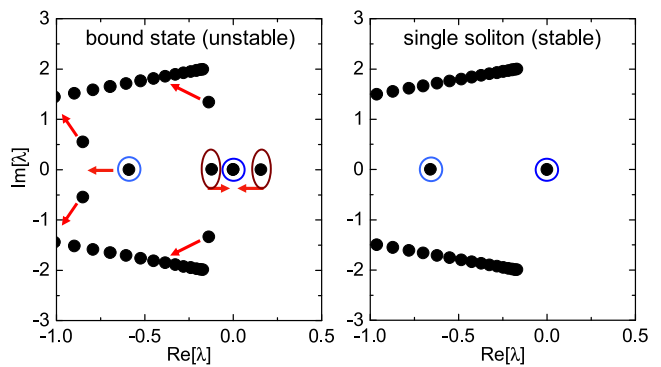} } \caption{(Color online) Complex eigenvalues for both the first level unstable
stationary in-phase BS and stable SP solution.  The arrows denote
the directions where the eigenvalues are shifted for higher levels
BSs with decreasing interaction; model: CQGLE (7); parameters:
$k=0.5$, $\beta=1$.}
\end{figure}

Evidently, in the limit of two infinitely separated SPs the
eigenvalues coincide with those of the stable single SP solution. In
this case the zero eigenvalue exhibits a fourth order degeneracy
corresponding to the two phase and two translation modes. If the
peak separation decreases the pulses start to perturb each other
affecting the eigenvalues. Because of interaction two eigenvalues
branch off from the zero eigenvalue and move in opposite directions
on the real axis for decreasing separation. Thus, the eigenvalue
with the positive real part is responsible for the instability of
the BS solution, see Fig.\ 5. The two eigenmodes corresponding to
these eigenvalues are shown in Fig.\ 6. They are formed as linear
combinations of the four individual eigenmodes of the original SP
(two phase modes, two translation modes) shifted by the distance
$S$. The peaks of both modes are shifted towards the center and do
not coincide with the peaks of the BS. Also because of SPs
interaction another four eigenvalues are symmetrically split off the
spectral continuum and move away from it for decreasing BS level.
However, all of them have a negative real part and do not affect the
stability of the bound state solution.

\begin{figure}
\centerline{
\includegraphics[width=8.5cm, clip]{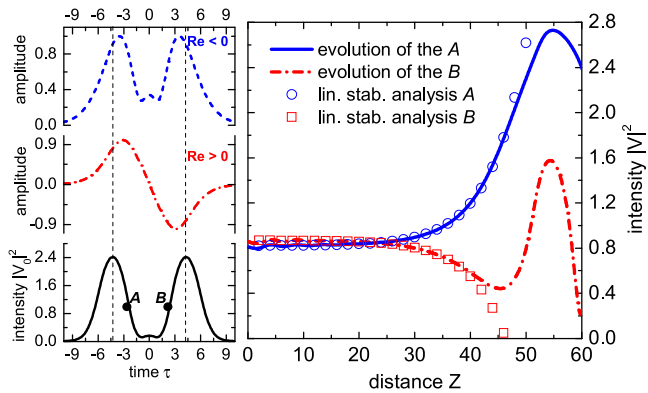} } \caption{(Color online) Left: Perturbation modes
(dotted blue line and dashed red line) of the first level in-phase
BS (solid black line). Right: Evolution trajectories of points
\textit{A} and \textit{B} of the stationary BS solution. The
amplitude of the perturbation modes is given in arbitrary units;
model: CQGLE (7); parameters: $k=0.5$, $\beta=1$.}
\end{figure}

From the results presented in Figs.\ 5 and 6 we may conclude that
the antisymmetric mode is responsible for the BS instability. In the
propagation simulations this mode is excited by noise and then grows
with $\mathrm{Re}(\lambda)>0$. To confirm the results of the linear
stability analysis we add white noise to the stationary BS solution
and plot the evolution trajectories of two points (fixed on the time
coordinate, calculated in the propagation scheme). The points are
chosen to be close to the amplitude maxima of the unstable mode, see
Fig.\ 6 (left). The growing antisymmetric eigenmode evokes an
increase of the amplitude in one point and a decay in the other
point upon propagation. Figure 6 (right) shows that the results of
the linear stability analysis and the exact numerical propagation
simulations perfectly coincide up to the distance where the linear
stability analysis ceases to be valid.

As expected, $\mathrm{Re}(\lambda)$ calculated for other BS levels
is smaller for higher levels, see Table \ref{t1}. This dependency
correlates well with the slower dynamics (evolution to the BS
solution and instability dynamics) discussed in Fig.\ 1. From a
mathematical and practical point of view the growth rate
$\mathrm{Re}(\lambda)$ for higher BS levels can be very close to
zero, but physically it remains an unstable BS.

\setlength{\tabcolsep}{1mm} \begin{table}
\caption{$\mathrm{Re}(\lambda)$ for different levels of stationary
BS solutions; model: CQGLE (7); parameters: $k=0.5$, $\beta=1$.}
\label{t1}
\begin{tabular}{c|c|c|c|c}
 & \multicolumn{2}{c|}{in-phase} & \multicolumn{2}{c}{out-of-phase} \\
level & peak & $\mathrm{Re}(\lambda)$ & peak & $\mathrm{Re}(\lambda)$ \\
 & separation &  & separation &  \\
\hline
1 & 8.5  & $1.57\cdot 10^{-1}$  & 4.7  & $5.95\cdot 10^{-1}$ \\
2 & 14.5 & $4.21\cdot 10^{-3}$  & 11.7 & $3.11\cdot 10^{-2}$ \\
3 & 20.3 & $6.09\cdot 10^{-5}$  & 17.3 & $5.11\cdot 10^{-4}$ \\
4 & 25.9 & $8.60\cdot 10^{-7}$  & 23.1 & $7.23\cdot 10^{-6}$ \\
5 & 31.7 & $1.17\cdot 10^{-8}$  & 28.7 & $1.02\cdot 10^{-7}$ \\
6 & 37.7 & $1.13\cdot 10^{-10}$ & 34.7 & $6.10\cdot 10^{-10}$ \\
\end{tabular}
\end{table}

We varied the CQGLE parameters $\beta$, $k$ and $\delta$ to find
in-phase or out-of-phase stable BSs but failed.

\section{\label{sec:level1}The lumped model}

In this section we study BS formation in fiber ring lasers using the more
realistic lumped model and experimentally accessible laser parameters. The aim
of this study is manifold. First of all, this model reflects better the
experimental situation. So, it is interesting to compare the results with those
of the distributed model which is easier to handle but contains severe
approximations. Moreover, we will study the effect of the finite SA response
time (noninstantaneous model) on the BS stability. A stabilization has been
proven in the spatial domain by taking advantage of nonlocality (see \cite{32}
and references therein). Here, the noninstantaneous response plays the role of
nonlocality in the spatial domain. Eventually we will summarize the differences
between the results provided by the lumped and the distributed model in order
to derive the limits of validity of the latter.

For the simulation we use parameters closely related to previous experiments at
a carrier wavelength of 1030 nm \cite{22, 23, 27}. The length of the absorber
and the small-signal loss were adjusted to a modulation depth of 30\%, the
relaxation time of the fast, but noninstantaneous, SA is 500 fs, its saturation
energy amounts to 16.7 pJ and the respective saturation power to
$P_\mathrm{sat}=33.4$ W. The output loss is equal to 30\%. For the doped fiber
we assume $L_\mathrm{f}=1$ m, $\beta_2=0.024 ~ \mathrm{ps}^{2}\mathrm{m}^{-1}$,
$\gamma=0.005 ~ \mathrm{W}^{-1}\mathrm{m}^{-1}$ and
$E^\mathrm{Gain}_\mathrm{sat}=1$ nJ. Amplitude and phase profiles were always
recorded after the output coupler.

In order to trigger BS formation we varied the filter strength, corresponding
to the inverse gain bandwidth $T_1$ of the laser material, and the small-signal
gain.

\subsection{\label{sec:level2}Instantaneous SA approximation}

In a first step we aim at reproducing the BS levels obtained in the
CQGLE model. For this purpose we start with the instantaneous SA
approximation, Eq.\ (4). As initial condition we use two in-phase or
out-of-phase resting small amplitude Gaussian pulses which converge
to the BS. Because of the additional stabilization effect of the
saturated gain the current model is less critical with respect to
the initial pulse profiles than the CQGLE. We observe BS formation
for $T_1=300$ fs and $g_0=0.75 ~ \mathrm{m}^{-1}$.

The first and second stationary in-phase BS levels are displayed in
Fig.\ 7. In-phase BSs exhibit a small characteristic central peak in
the intensity profile and the difference between phase maximum and
minimum (at the center) is a multiple of $\sim \pi$. In agreement
with our previous results the formation of the second level in-phase
BS needs approximately 80 times more (40,000) round trips than the
1st level (500). However, also these stationary BS solutions are
unstable with an asymmetric instability behavior which is similar to
the CQGLE BSs, see Fig.\ 8.

\begin{figure}
\centerline{
\includegraphics[width=8.5cm, clip]{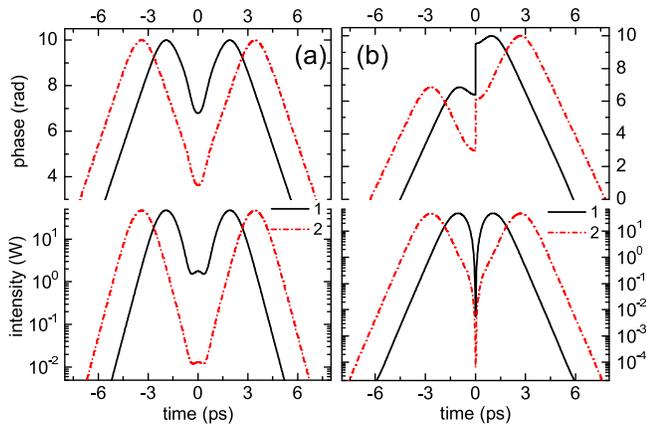} }\caption{(Color online) Intensity and phase profiles for first (solid line 1) and second (dashed line 2)
levels of the stationary BS solutions; (a) - in-phase, (b) -
out-of-phase; lumped model (parameters, see text), instantaneous SA
approximation (4).}
\end{figure}

\begin{figure}
\centerline{
\includegraphics[width=6cm, clip]{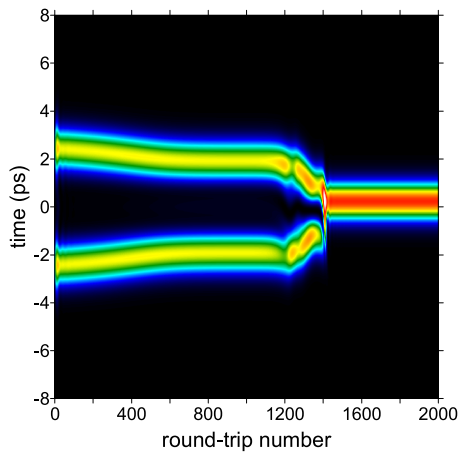} }\caption{(Color online) Bound state instability. The two in-phase pulses form the stationary bound state and fuse asymmetrically eventually; lumped model (parameters, see text), instantaneous SA approximation (4).}
\end{figure}

For out-of-phase initial conditions we  have also observed first and
second level stationary out-of-phase BSs, see Fig.\ 7. They exhibit
a similar formation dynamics and instability behavior as in-phase
solutions. In particular, the formation of a level 2 out-of-phase BS
needs approximately 30 times more (3,000) round-trips than that for
a level 1 out-of-phase BS (100).

To visualize the interactions between both pulses we again take
advantage of the phase plane. We consider the evolution trajectories
for initially in-phase and out-of-phase pulses, see Fig.\ 9. For
in-phase pulses the trajectory is partly described by four circles.
As we already know, each circle $(R_1,~R_2,~R_3,~R_4)$ corresponds
to a certain level of out-of-phase or in-phase stationary BS
solutions, shown in Fig.\ 9. There is essentially no difference
between the evolution trajectories provided by the lumped and the
distributed model.

\begin{figure}
\centerline{
\includegraphics[width=8.5cm, clip]{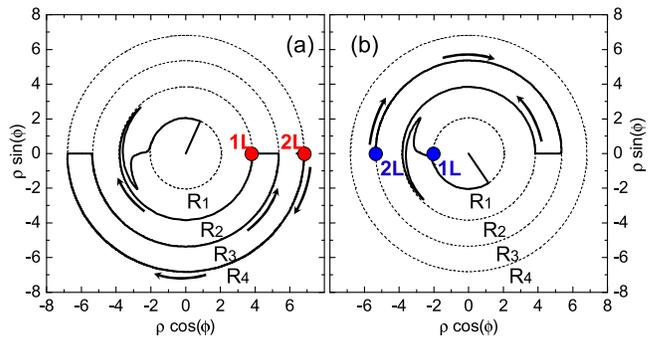} }\caption{(Color online) Pulse trajectories in
the phase plane for two pulses (a) initially in-phase with separation 6.38 ps
and (b) initially out-of-phase with separation 5.25 ps. 1L - first level BS, 2L
- second level BS; lumped model (parameters, see text), instantaneous SA
approximation (4).}
\end{figure}

\subsection{\label{sec:level2}Differences between models}

For the sake of comparison between the different models and of evaluating their
validity we use the instantaneous SA approximation and compare results provided
by the lumped model, which serves as a benchmark, with both distributed models,
the MGLE (5), which accounts for the full SA saturation, and the CQGLE (7),
which relies on the Taylor expansion of the SA saturation.

Figure 10 shows stable SP solutions obtained by both the lumped and
the MGLE model. For the lumped modeling the saturation energy of the
gain was reduced by a factor of two, because we consider just a
single pulse. The MGLE parameters were calculated from those of the
lumped model, except that for the distributed model a proper
magnitude of the small-signal gain (6) was adjusted as  $g_0 =
0.6388 ~ \mathrm{m}^{-1}$, corresponding to the gain in the lumped
model for stationary conditions. It is interesting to note that for
the relevant parameter set stable SP solutions exist in the MGLE,
but not in the CQGLE model. This is a clear indication that the
saturation behavior plays a pivotal role. Usually mode-locked lasers
are working in a regime where the absorber is saturated, i.e.,
$|V|^2/P^\mathrm{aver}_\mathrm{sat}<1$ is not satisfied and the
CQGLE is incorrect. Hence, one must not expect solutions for similar
parameters in the CQGLE model on the one side and the lumped (or
MGLE) model on the other side. Nevertheless, the MGLE and the lumped
model with an ideal SA show quite a good agreement regarding the
stable SP solutions, see Fig. 10.

\begin{figure}
\centerline{
\includegraphics[width=8.5cm, clip]{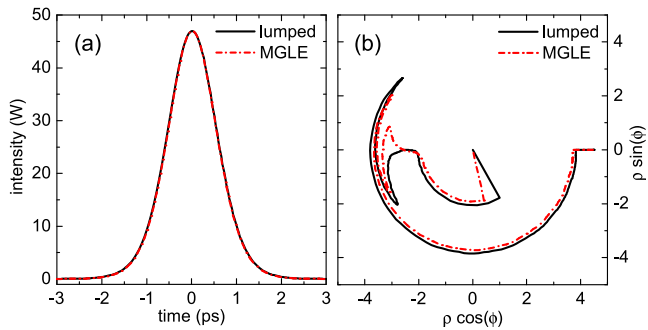} }\caption{(Color online) Comparison between
the lumped and the MGLE model at the same parameters set. (a) - intensity profile of stable
SP solutions, (b) - evolution trajectories of two pulses in the phase plane; instantaneous SA
approximation (4).}
\end{figure}

Further on we consider the evolution of two SPs in the phase plane, see Fig. 10
(in this case we use the original value of the saturation energy
$E^\mathrm{Gain}_\mathrm{sat} = 1$ nJ). Although minor differences occur the
evolution trajectories of the two pulses behave quite similar for both the MGLE
and the lumped model.

\subsection{\label{sec:level2}Noninstantaneous SA response}

In a next step, relying on the lumped model, we are attempting to
achieve BS stabilization by accounting for the noninstantaneous SA
response (3). This model reflects the experimental situation because
typical SA response times vary between 300 fs and 12 ps, which
frequently compares to the pulse length. We use the parameters from
the previous section \textbf{A}, in order to keep the saturation
power constant, and account for the variation of the relaxation time
by a proper change of the saturation energy
$E^\mathrm{SA}_\mathrm{sat}$.

In these simulations as initial condition we use two resting pulses with small
amplitudes. At the beginning the pulses propagate 100 round trips in the ideal
SA approximation and just then we switch on the noninstantaneous response. This
procedure is required to arrive at a definite initial condition for large
amplitude bound states.

Trivially, for a vanishing relaxation time the system exhibits
unstable stationary BS solutions identical to them of the
instantaneous SA approximation. With increasing relaxation time
(about 150 fs) we observe a tendency towards BS stabilization with
moderate oscillations. Eventually the BSs stabilize for a relaxation
time (about 350 fs) close to the experimental situation, see Figs.\
11, 12. With a further increase of the relaxation time the arising
stable BSs exhibit a fixed peak separation. The solutions move
backwards in the reference frame due to the temporal effects in the
absorber, see Fig.\ 12.

\begin{figure}
\centerline{
\includegraphics[width=8.5cm, clip]{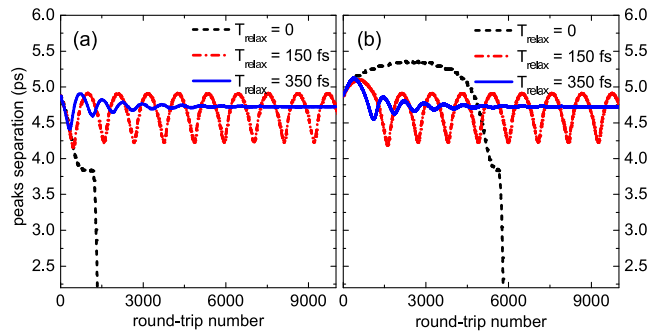} }\caption{(Color online) Bound state stabilization.
Distance between two pulses as a function of round-trips for initially (a) - in-phase and (b) -
out-of-phase pulses for different SA relaxation times. In the limit $T_\mathrm{relax}=0$,
pulses build (a) the level 1 in-phase stationary BS solution, (b) the level 2 out-of-phase
stationary BS solution; lumped model (parameters, see text), noninstantaneous SA response (3).}
\end{figure}

\begin{figure}
\centerline{
\includegraphics[width=6cm, clip]{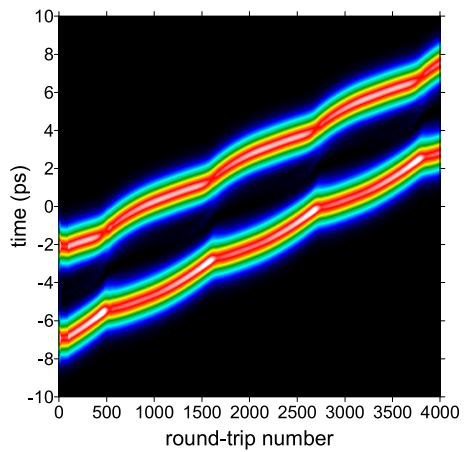} }\caption{(Color online) Oscillating BS solution for $T_\mathrm{relax}=150$ fs; lumped model (parameters, see text), noninstantaneous SA response (3).}
\end{figure}

It is interesting to note that the resulting stationary and
oscillating stable BS solutions do not depend on the initial phase
difference between the pulses, see Fig.\ 11. This is in stark
contrast to the CQGLE model, where the initial phase difference
defines the nature of the stationary BS solution. There, two
in-phase or out-of-phase pulses with identical separation evolve
towards different levels (see the curve for $T_\mathrm{relax} = 0$
in Fig.\ 11). Moreover, we have observed stationary and oscillating
stable BS solutions of the next level with a larger peak separation.
To obtain them the initial distance between pulses was increased and
proper figures for the relaxation time were chosen. Higher level BSs
need approximately 100 times more round-trips for their formation,
which is in agreement with previous results.

The nature of stationary and oscillating stable BS solutions can be
understood by inspecting the trajectories in the phase plane, see
Fig.\ 13. We evaluate the two pulse dynamics in phase space for four
different cases. To generate the first and second level BSs we use
two pulses with an initial separation of 4.9 ps and 6 ps,
respectively. Moreover, for each case we adjust the proper
relaxation time in order to obtain a stationary or an oscillating
stable BS solution. Figure 13 shows that stationary BS solutions are
fixed points in the phase plane which are located near $\phi=\pi/2$.
Oscillating solutions move either in the lower or upper half plane
between two adjacent circles.

\begin{figure}
\centerline{
\includegraphics[width=8.5cm, clip]{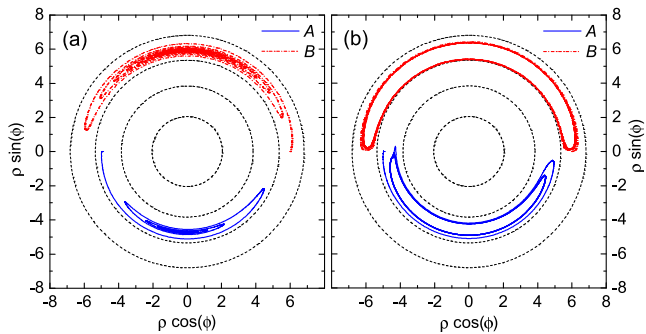} }\caption{(Color online) Trajectories showing the
two-pulse evolution; a) 1st level stable BS solution
$T_\mathrm{relax}=350$ fs (blue line \textit{A}), 2nd level stable
BS solution $T_\mathrm{relax}=75$ fs (red line \textit{B}) and (b):
1st level oscillating BS solution $T_\mathrm{relax}=150$ fs (blue
line \textit{A}), 2nd level oscillating BS solution
$T_\mathrm{relax}=5$ fs (red line \textit{B}); lumped model
(parameters, see text), noninstantaneous SA response (3).}
\end{figure}

The conclusion to be drawn is that stationary stable BS solutions
(Fig.\ 14) of the lumped model with noninstantaneous SA are neither
in- nor out-of-phase but exhibit a phase difference of about $\pi/2$
between the pulses. Immediately the question arises whether there
are stable BS solutions with $\pi/2$ phase difference between the
pulses in the instantaneous models (CQGLE or lumped model with
instantaneous SA). In the lumped model no stable or even stationary
BS solutions with that phase difference could be identified. For the
CQGLE model, but just for anomalous dispersion, this case has been
studied earlier \cite{21}, and it was shown, that stable BS
solutions can exist.

\begin{figure}
\centerline{
\includegraphics[width=6cm, clip]{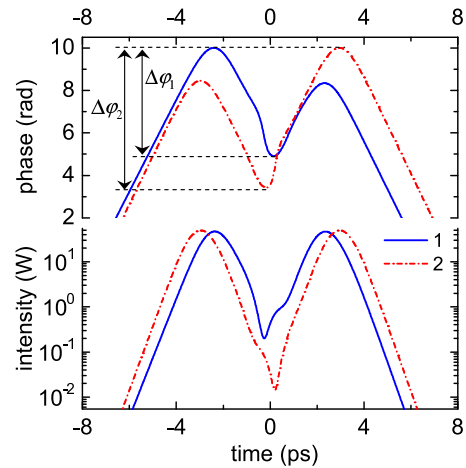} }\caption{(Color online) The intensity and
phase profiles for 1st and 2nd level stable BS solutions calculated for
$T_\mathrm{relax}=350$ fs (blue solid line 1), $T_\mathrm{relax}=75$ fs (red
dashed line 2). The phase differences are $\Delta\varphi_1=5.1$ and
$\Delta\varphi_2=6.6$; lumped model (parameters, see text), noninstantaneous SA
response (3).}
\end{figure}

Nevertheless, stable BS solutions in the lumped model with
noninstananeous SA differ considerably from those discussed in
\cite{21}. First, they exist for a completely different set of
parameters (even opposite dispersion) and second, their behavior in
phase space (Fig.\ 13) exhibits a large difference to those studied
in \cite{21}. Because all SAs have a finite relaxation time the
conclusion can be drawn that the parameter range for the existence
of BSs is much wider than assumed in \cite{21}.

As already mentioned BS stabilization is evoked by the noninstantaneous SA
response in analogy to the spatial case where a nonlocality stabilizes
localized solutions \cite{32}. Here, the retarded temporal SA response is
asymmetric in time, leading to the minor deviation of the fixed point from
$\phi=\pi/2$. By contrast, the stable BS solutions of the CQGLE appear exactly
for $\phi=\pi/2$ \cite{21}.

The region of existence of stable BSs is displayed in Fig.\ 15. It
can be recognized that the regions for two neighboring BS levels
differ, especially the bottom boundaries. This means that
oscillating solutions of the first and second level can be observed
for different relaxation times and the same small-signal gain. The
open boundary indicates that the bottom margin is very close to the
zero axis.

\begin{figure}
\centerline{
\includegraphics[width=8.5cm, clip]{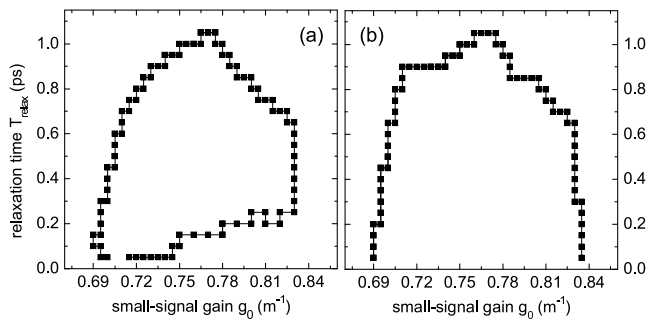} }\caption{Existence domains for
stable (stationary and oscillating) bound state solutions; (a) first level solution,
(b) second level solution; lumped model (parameters, see text), noninstantaneous SA response (3).}
\end{figure}

A distinctive feature of stable BS solutions is their behavior in
the phase plane. The two initially resting pulses, for a certain
range of initial separation, always evolve towards a single fixed
point independently of the initial phase difference. This is in
stark contrast to the stable BS solutions studied in Ref. \cite{21}
where two similar solutions are possible, lying symmetrically on
opposite sides of the phase plane. Moreover, the neighboring levels
of BSs are situated in different half spaces, see Fig.\ 13. This is
due to the lack of invariance of the time scale if relaxation
matters. Thus leading and trailing pulses are not equivalent.

Figure 13 shows that each fixed point on the phase plane lies
between two neighboring circles. Thus one might anticipate that
there is a fixed point between circle 1 and 2 too. But in our
simulations we did not succeed in finding any stable BS in that
domain.

\section{\label{sec:level1}Conclusion}

In conclusion, we have theoretically and numerically investigated the formation
and stability of BSs in a mode-locked fiber ring laser with SA in the normal
dispersion regime. We started our investigations with the distributed
description of the laser which is given by the commonly used CQGLE. For the
first time, to the best of our knowledge, we have observed a discrete family of
stationary BS solutions with different peak separation. Moreover, the observed
BS levels exhibit an equidistantly increasing peak separation with a
well-defined phase relation. All levels of in-phase and out-of-phase BSs turned
out to be unstable for the investigated system parameters. A linear stability
analysis has shown that this instability is caused by the antisymmetric
perturbation mode. We found that both the formation dynamics of BSs and their
succeeding decay is slower for higher levels because of the weaker interaction.
The evolution trajectories in the phase plane of two SPs with relatively large
initial separation consist of simple geometrical forms.

All these findings could be confirmed by using a more realistic lumped laser
model but maintaining the instantaneous SA response.

Stabilization of the BSs was achieved in the lumped laser model by taking the
fast, nevertheless noninstantaneous SA response into account. Using the SA
relaxation time as control parameter a continuous transition from unstable to
stable BS solutions may be achieved passing a domain of oscillating BS
solutions. These stationary and oscillating stable BSs have a discrete
multilevel nature too. The distinctive feature of these BS solutions is
expressed in the phase plane. A stable BS solution of a certain level
represents a fixed point in the phase plane, which lies in the prefered half
space. For oscillating BSs the evolution is described by circulations between
two neighboring circles in either the lower or upper half space depending on
the level. The stabilization mechanism identified shows that all lasers with
semiconductor absorber have a wider region of stable BS solutions than
previously assumed \cite{21}. Moreover, we have shown that a discrete family of
stable or oscillating BS solutions with equidistant peak separation can be
obtained in such laser systems.

The results presented can be used in optics communication lines for generating
multipulse trains with well defined separation distances or in fiber lasers for
an irregular or regular pulse array generation.

\begin{acknowledgments}
This work was supported by the Deutsche Forschungsgemeinschaft
(research unit 532). Rumen Iliew acknowledges financial support from
the Carl Zeiss foundation.
\end{acknowledgments}

%\bibliographystyle{plain}
%\bibliography{apssamp_1}% Produces the bibliography via BibTeX.

\end{document}